\documentclass[conference]{IEEEtran}
\IEEEoverridecommandlockouts
\usepackage{cite}
\usepackage{amsmath,amssymb,amsfonts}
\usepackage{algorithm}
\usepackage{algpseudocode}
\usepackage{graphicx}
\usepackage{textcomp}
\usepackage{xcolor}
\usepackage{subfig}
\usepackage{booktabs}
\usepackage{import}
\usepackage{color}
\usepackage[citebordercolor=blue]{hyperref}
\usepackage[capitalise]{cleveref}
\usepackage{multirow}
\usepackage{array}
\usepackage[font=footnotesize]{caption}
\usepackage[inline]{enumitem}
\usepackage{makecell}
\usepackage{siunitx}
\usepackage{caption}

\usepackage{tikz}
\newcommand\copyrighttext{%
  \footnotesize \textcopyright 2026 IEEE. Personal use of this material is permitted. Permission from IEEE must be obtained for all other uses, in any current or future media, including reprinting/republishing this material for advertising or promotional purposes, creating new collective works, for resale or redistribution to servers or lists, or reuse of any copyrighted component of this work in other works.}
\newcommand\copyrightnotice{%
\begin{tikzpicture}[remember picture,overlay]
\node[anchor=south,yshift=10pt] at (current page.south) 
  {\fbox{\parbox{\dimexpr\textwidth-\fboxsep-\fboxrule\relax}{\copyrighttext}}};
\end{tikzpicture}%
}

\begin{document}
\bstctlcite{BSTcontrol}
%
\title{Location-Aware NAS Timer Optimization in NTN-TN Integrated Networks}
%
%
%

\author{\IEEEauthorblockN{
    Cheng~Liu~and~Peng~Hu\textsuperscript{*}
\IEEEauthorblockA{
    Advanced Network and Embedded Systems Lab (AEL)\\
    Dept. of Electrical and Computer Engineering, University of Manitoba, Winnipeg, Canada}
    {liuc13@myumanitoba.ca, peng.hu@umanitoba.ca}
}

\thanks{We acknowledge the support provided by the Government of Canada, and the Natural Sciences and Engineering Research Council of Canada (NSERC), [funding reference number RGPIN-2022-03364].}
}

%
%

\markboth{Journal of \LaTeX\ Class Files,~Vol.~14, No.~8, August~2015}%
{Shell \MakeLowercase{\textit{et al.}}: Bare Demo of IEEEtran.cls for IEEE Journals}
%



\maketitle
\copyrightnotice
\begin{abstract}

Efficient Non-Access Stratum (NAS) timer configuration is critical for reliable and energy-efficient Fifth Generation (5G) registration in Non-Terrestrial Network (NTN)-Terrestrial Network (TN) integrated systems, where Low Earth Orbit (LEO) satellite access introduces large registration bursts, heterogeneous propagation paths, and multi-hop satellite routing. Existing 3GPP NAS timers use fixed values, while prior closed-form timer models compute a global timer under network-level assumptions; both fail to capture user equipment (UE)-level differences in propagation delay, Access and Mobility Management Function (AMF) arrival position, and path reliability. In this paper, we propose a location-aware, UE-specific NAS timer optimization method for LEO NTN-TN integrated networks. The proposed method models the path-delay component using service-link geometry, ground-station distance, and Inter-Satellite Link (ISL) hop count, and adapts the endpoint-delay component according to each UE's expected AMF queue exposure and path reliability. Simulation results show that our method reduces registration latency, UE energy consumption, and avoidable registration attempts compared with fixed and global timer configurations, especially when timer over-provisioning causes unnecessary waiting.

\end{abstract}

\begin{IEEEkeywords}
Low Earth Orbit satellite networks, Non-Terrestrial Networks, Non-Access Stratum, 5G registration, Timer adaptation.
\end{IEEEkeywords}

%
\IEEEpeerreviewmaketitle

\section{Introduction}

Non-Terrestrial Networks (NTNs) are becoming an important component of future Fifth Generation (5G) and Sixth Generation (6G) systems \cite{araniti2021toward}, especially for extending service coverage to remote, maritime, aerial, and disaster-affected areas. In NTN-Terrestrial Network (TN) integrated systems, user equipments (UEs) may access the 5G core through Low Earth Orbit (LEO) satellites, ground stations (GSs), and terrestrial core network functions \cite{3gpp_tr38821}. Before receiving network services, a UE must complete the Non-Access Stratum (NAS) registration procedure with the Access and Mobility Management Function (AMF) \cite{3gpp_ts24501}. The efficiency of this procedure directly affects access latency, signaling overhead, and UE energy consumption.

NAS timers play a key role in NTN-TN integrated 5G-and-beyond systems. During registration, timers such as T3510, T3550, and T3560 determine how long the UE or AMF waits for the expected NAS response before retransmission. If a timer is too short, delayed but valid responses may be treated as failures, causing unnecessary retransmissions and additional signaling load. If a timer is too long, the UE may wait excessively before detecting packet loss or procedure failure, increasing registration delay and energy consumption. This tradeoff is particularly challenging in LEO NTN-TN integrated networks because propagation delay, inter-satellite routing, access bursts, and packet loss vary significantly across UEs \cite{soret2024q}.

Existing 3rd Generation Partnership Project (3GPP) NAS timer values are fixed and simple to deploy, but they do not account for NTN-specific path heterogeneity. A UE close to a GS may reach the core network with few or no Inter-Satellite Link (ISL) hops, while another UE may traverse multiple ISLs before its NAS message exits the satellite constellation \cite{wu2025enhancing}. As a result, a single fixed timer may be overly conservative for short-path UEs and insufficient for long-path or queue-exposed UEs. Earlier work on NTN layer-2 (L2) timers has shown that round-trip variability requires re-evaluating fixed configurations \cite{sheemar2023adaptive}. More recently, closed-form NAS timer models have been proposed for LEO constellations by incorporating link variability and processing delay \cite{hezaveh2026astrotimer}. However, when applied as global timer configurations, they still cannot fully capture UE-level differences in satellite path length, AMF arrival position, and path reliability.

In this paper, we propose a location-aware NAS timer optimization method for NTN-TN integrated networks. The key idea is to preserve the efficiency of closed-form timer computation while making the timer value UE-specific. We model the path-delay component using service-link (SL) geometry and ISL hop count, so UEs with different satellite-to-GS paths receive different propagation budgets. We further adapt the endpoint-delay component according to each UE's expected AMF queue exposure \cite{alawe2018scalability} and path reliability, allowing the timer to become more conservative only when a UE is likely to benefit from additional waiting.

The main contributions of this paper are summarized as follows:
\begin{itemize}
    \item We formulate UE-specific NAS timer sizing for LEO NTN-TN integrated networks by modeling the path-delay component from SL geometry, GS distance, and ISL hop count.
    \item We design an adaptive endpoint-weighting mechanism that accounts for each UE's expected AMF queue exposure and path reliability, reducing unnecessary timer over-provisioning while avoiding premature retransmissions.
    \item We evaluate the proposed method through end-to-end NAS registration simulations and show that it reduces registration latency, UE energy consumption, and avoidable registration attempts compared with fixed and global timer configurations.
\end{itemize}
The rest of the paper is structured as follows: Section II discusses the related work; Section III discusses the proposed methodology; Section IV presents the experimental results; and Section V concludes the paper and outlines the future work. 

\section{Related Work}

The integration of NTNs into 5G and beyond systems has attracted substantial attention as a means of delivering ubiquitous connectivity beyond the reach of terrestrial infrastructure \cite{araniti2021toward,wang2025non,jamshed2025tutorial}. Architectural studies have explored how LEO constellations can be coupled with the 5G core through transparent or regenerative payloads, ISLs, and ground gateways \cite{3gpp_tr38821,yuan2025satellite,di2018ultra}, highlighting that satellite access fundamentally differs from terrestrial access in propagation delay, link variability, and topological dynamics. To accommodate these effects, routing works on LEO constellations have investigated how hop count, link length, and queueing at intermediate satellites jointly shape end-to-end latency \cite{wu2025enhancing,rabjerg2021exploiting,soret2024q}, showing that the satellite path traversed by a packet is highly UE-dependent rather than a single network-wide quantity. Mobility management studies further report that LEO geometry induces frequent handovers and signaling bursts that stress the control plane \cite{zhang2024secure,juan2022handover}. These observations motivate revisiting control-plane procedures that were originally designed under terrestrial timing assumptions.

Within the 5G control plane, NAS registration is the entry-point procedure that governs UE access latency, signaling overhead, and energy consumption \cite{3gpp_ts24501}. A series of studies has addressed the resulting signaling pressure from the core-network side: Alawe~\textit{et~al.} \cite{alawe2018scalability} modeled the AMF as a queuing system and proposed control-theoretic load balancing and scale-out/scale-in policies to absorb registration bursts, while subsequent studies advocate stateless or distributed core-network designs for satellite so that registration signaling can be processed closer to the access edge \cite{liu2023stateless}. Other proposals target the radio side, mitigating registration congestion through random-access preamble design and access-class-barring schemes adapted to NTN propagation \cite{chougrani2022nbiot}. These efforts focus on provisioning sufficient capacity for the signaling that does arrive at the AMF; however, the duration each UE waits for a NAS response before declaring failure is still controlled by NAS watchdog timers, which remain largely outside the scope of these studies. As a result, the interaction between the AMF queue formed under bursty access and the UE-side waiting policy is not directly addressed.

Recent studies have examined timer configurations. The 3GPP NAS specification provides only scaled fixed values for satellite access, derived from MEO/GEO assumptions and inherited from terrestrial defaults; these values are conservative and uniform across UEs, and the standard explicitly leaves LEO-specific timer settings undefined \cite{3gpp_ts24501}. At L2, Sheemar~\textit{et~al.} \cite{sheemar2023adaptive} have shown that packet data convergence protocol (PDCP) discard, PDCP reordering, and radio link control (RLC) reassembly timers can be tightened in NTN by estimating the effective number of retransmissions rather than assuming the maximum numbers of lower-layer hybrid automatic repeat request (HARQ) and automatic repeat request (ARQ) retransmissions, improving throughput and buffer occupancy. While this confirms that fixed timers are wasteful under NTN propagation, L2 adjustments do not translate to the NAS layer, which spans the entire UE--AMF path and is exposed to ISL multi-hop routing and core-network queueing. Most recently, AstroTimer \cite{hezaveh2026astrotimer} introduced a closed-form NAS timer model for LEO constellations that outperforms 3GPP defaults by accounting for link variability, processing delays, and network-function placement. However, it computes a single global timer, failing to distinguish between UEs on short satellite paths and those traversing multiple ISL hops with AMF queuing delays. This global approach also ignores per-UE path reliability, potentially forcing UEs to wait unnecessarily for messages already lost in transit. To overcome the limitations of global timers, we formulate a UE-specific, location-aware NAS timer that jointly captures SL geometry, ISL hops, AMF queue exposure, and path reliability. This tailored approach mitigates registration inefficiencies while maintaining a lightweight, closed-form structure deployable in operational NTN-TN integrated networks.

\section{Proposed Methodology}
In this section, we present the system model and the proposed location-aware NAS registration timer policy. 

\begin{figure}[!t]
\centering
\includegraphics[width=1\linewidth]{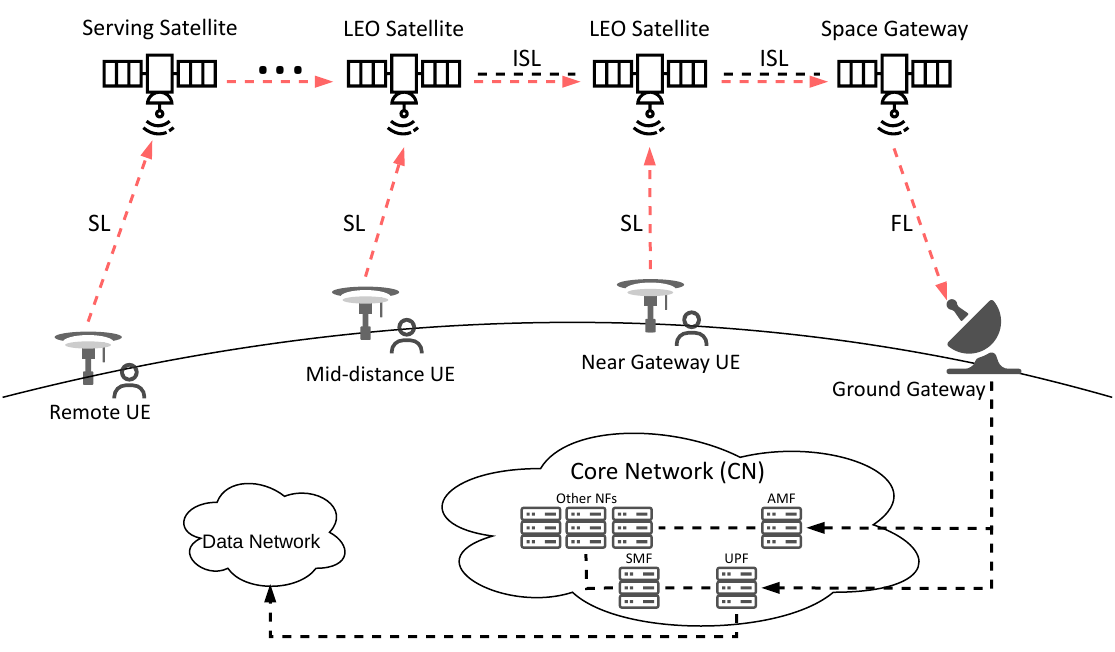}
\caption{Architecture of an LEO satellite network.}
\label{fig:arch}
\end{figure}

\subsection{Background and Problem Statement}

We consider a 3GPP-compliant NTN-TN integrated network consisting of UEs on the Earth surface, a Walker-delta LEO satellite constellation, GSs, and 5G core network functions. As shown in Fig.~\ref{fig:arch}, during NAS registration, signaling messages are delivered from a UE to its serving satellite, forwarded through zero or more ISLs, relayed through a GS and Feeder Link (FL), and finally processed by the AMF. The registration procedure is governed by NAS watchdog timers, including T3510 at the UE and T3550/T3560 at the AMF. Timer expiration triggers retransmission and may increase signaling load and UE energy consumption.

We build on the closed-form NAS timer formulation introduced in AstroTimer~\cite{hezaveh2026astrotimer}, which decomposes a timer into a path-delay component and an endpoint-delay component:
\[
T = R P + \gamma E ,
\]
where $R$ is the number of NAS message exchanges, $P$ is the one-way path delay between the timer origin and responder, $E$ is the weighted endpoint delay, and $\gamma=\lfloor R/2 \rfloor + 1$ accounts for how endpoint delays are accumulated across the NAS exchange.

The path-delay component is defined as
\[
P =
D_{\mathrm{prp}}^{\langle 0,1\rangle}
+
\sum_{i=1}^{N-1}
\left(
D_{\mathrm{agg}}^{i}
+
D_{\mathrm{prp}}^{\langle i,i+1\rangle}
\right)
+
D_{\mathrm{prp}}^{\langle N-1,N\rangle},
\]
where $D_{\mathrm{prp}}^{\langle i,i+1\rangle}$ is the propagation delay between adjacent nodes, $D_{\mathrm{agg}}^{i}$ is the aggregated processing and queueing delay at the intermediate node $i$, and $N$ is the number of hops between the timer origin and responder.

The endpoint-delay component is given by
\[
E = \alpha D_{\mathrm{agg}}^{0} + \beta D_{\mathrm{agg}}^{N},
\]
where $D_{\mathrm{agg}}^{0}$ and $D_{\mathrm{agg}}^{N}$ denote the aggregated delay at the timer origin and responder, respectively, and $\alpha,\beta \in (0,1]$ are endpoint weighting coefficients.

While this closed-form formulation captures multi-hop propagation and burst-induced queueing delay, applying it with fixed path assumptions and fixed endpoint weights is insufficient for a large-scale LEO registration scenario. In this sense, UEs at different geographic locations may attach to different serving satellites and traverse different numbers of ISL hops before reaching a GS. Therefore, the path-delay component $P$ is inherently UE-dependent rather than a single network-wide quantity.

The endpoint-delay weight is also affected by LEO routing geometry. UEs with shorter satellite-to-GS paths reach the AMF earlier and contribute to the queue seen by later arrivals, while longer-path UEs arrive after both additional propagation and partial AMF service. A single global timer must therefore compromise between short-path and long-path UEs: it may expire prematurely for UEs with larger delay or queueing exposure, while over-provisioning UEs with shorter paths. This motivates a timer formulation that preserves the closed-form structure above while accounting for UE-specific path length, arrival position, and path reliability.

\subsection{UE-Specific Path Delay Modeling}

To account for spatial heterogeneity in LEO access, we first reformulate the path-delay component as a UE-specific quantity. For UE $u$, the one-way path from the UE to the AMF consists of an SL, a sequence of ISL hops, and an FL. We denote the resulting path-delay component by
\[
P(u) = D_{\mathrm{SL}}(u) + k_u \kappa D_{\mathrm{ISL}} + D_{\mathrm{FL}}
+ \sum_{m=1}^{k_u+1} D_{\mathrm{agg}}^{\mathrm{sat},m},
\]
where $D_{\mathrm{SL}}(u)$ is the SL propagation delay, $k_u$ is the number of ISL hops, $D_{\mathrm{ISL}}$ is the per-hop ISL propagation delay, $\kappa$ as a coefficient captures path inflation relative to the shortest geometric route, $D_{\mathrm{FL}}$ is the FL propagation delay, and $D_{\mathrm{agg}}^{\mathrm{sat},m}$ denotes the aggregated delay at the intermediate satellite nodes.

The SL delay depends on the UE's satellite elevation angle. Let $h$ be the satellite altitude and $c$ be the speed of light. We compute
\[
D_{\mathrm{SL}}(u) = \frac{d_{\mathrm{slant}}(\theta_{\mathrm{eff}}(u), h)}{c},
\]
where $d_{\mathrm{slant}}(\cdot)$ is the slant distance between the UE and its serving satellite. The effective elevation angle is defined as
\[
\theta_{\mathrm{eff}}(u)
=
(1-\eta)\theta_{\mathrm{mean}}(u)
+
\eta\theta_{\mathrm{min}}(u),
\]
where $\theta_{\mathrm{mean}}(u)$ and $\theta_{\mathrm{min}}(u)$ represent the average and minimum visible elevation conditions for UE $u$, and $\eta$ controls how conservatively the model accounts for low-elevation access.

Next, we estimate the ISL hop count from the UE's distance to the nearest GS. Let
\[
\psi(u) = \min_{g \in \mathcal{G}} d_{\mathrm{gc}}(u,g),
\]
where $\mathcal{G}$ is the set of GSs and $d_{\mathrm{gc}}(\cdot)$ is the great-circle distance. Given a GS coverage radius $r_{\mathrm{GS}}$ and an average inter-satellite spacing $d_{\mathrm{sat}}$, the ISL hop count is
\[
k_u =
\begin{cases}
0, & \psi(u) \le r_{\mathrm{GS}},\\
\left\lceil \dfrac{\psi(u)-r_{\mathrm{GS}}}{d_{\mathrm{sat}}} \right\rceil, & \psi(u) > r_{\mathrm{GS}}.
\end{cases}
\]

This formulation turns the path-delay term from a single network-level value into a UE-dependent quantity. UEs near a GS obtain small $k_u$ and short $P(u)$, while UEs far from any GS accumulate additional ISL propagation and intermediate-node delay. The resulting $P(u)$ is then used as the path component in the NAS timer formulation.

\subsection{Adaptive Endpoint Weighting}

After obtaining the UE-specific path-delay component $P(u)$, we adapt the endpoint-delay component to the queueing condition expected by each UE. For timer type $j$, the proposed NAS timer is
\[
T_j(u)
=
R_j P(u)
+
\gamma_j \alpha(u)
\left(
D_{\mathrm{agg}}^{0}
+
D_{\mathrm{agg}}^{N}
\right),
\]
where $R_j$ is the number of NAS message exchanges associated with timer $j$, $\gamma_j=\lfloor R_j/2 \rfloor+1$, and $\alpha(u)$ is a UE-specific endpoint weighting coefficient. For simplicity, the same UE-specific coefficient is applied to both endpoint-delay terms. Under bursty registration load, the AMF-side term dominates the endpoint delay, so the adaptation is primarily governed by the UE's expected AMF queue exposure.

The coefficient $\alpha(u)$ is designed to capture the queueing exposure of UE $u$ at the AMF. Since the dominant difference in AMF arrival time is induced by the ISL hop count, we approximate the relative arrival position of a UE by $k_u$. Let
\[
C(k_u)=|\{v \in \mathcal{U}: k_v \le k_u\}|
\]
be the number of UEs whose messages are expected to reach the AMF no later than UE $u$. During the time required for UE $u$ to traverse its ISL path, the AMF can process part of the earlier burst. Therefore, the remaining queue observed by UE $u$ is estimated as
\[
Q(k_u)
=
\max\left(0,\,
C(k_u)-\mu_{\mathrm{NF}} k_u D_{\mathrm{ISL}}
\right),
\]
where $\mu_{\mathrm{NF}}$ is the AMF service rate. We then normalize this queue estimate by the number of registering UEs:
\[
\alpha_{\mathrm{pos}}(u)
=
\min\left(1,\frac{Q(k_u)}{|\mathcal{U}|}\right).
\]

The factor $\alpha_{\mathrm{pos}}(u)$ increases when a UE is expected to encounter a larger residual AMF queue. However, queueing exposure alone should not always lead to a longer timer. In lossy LEO paths, a registration attempt may fail due to packet loss rather than late response arrival; in such cases, excessive timer extension only delays retransmission and failure detection. We therefore weight the queue-position factor by the reliability of the UE's path.

Let $p_{\mathrm{SL}}$, $p_{\mathrm{ISL}}$, and $p_{\mathrm{FL}}$ denote the loss probabilities of the SL, each ISL hop, and the FL. The probability that a one-way path from UE $u$ reaches the AMF without packet loss is
\[
p_{\mathrm{path}}(u)
=
(1-p_{\mathrm{SL}})
(1-p_{\mathrm{ISL}})^{k_u}
(1-p_{\mathrm{FL}}).
\]
Using a fixed reference exchange count $R_{\mathrm{ref}}=5$,
we define the reliability weight as
\[
s(u)=p_{\mathrm{path}}(u)^{2R_{\mathrm{ref}}}.
\]
This term provides a conservative measure of path reliability.

Finally, the adaptive endpoint coefficient is
\[
\alpha(u)=\alpha_0+
\max\{0,\alpha_{\mathrm{pos}}(u)-\alpha_0\}s(u).
\]
where $\alpha_0$ is a conservative lower bound. This coefficient has two useful implications. First, since $s(u)\in[0,1]$, the endpoint weight is always lower-bounded by $\alpha_0$, preventing the timer from becoming more aggressive than the chosen baseline compensation. Second, as path reliability decreases, $s(u)$ decreases and $\alpha(u)$ moves toward $\alpha_0$; therefore, the timer is not extended simply because a path is long when the registration attempt is unlikely to succeed due to packet loss. Additional conservativeness is applied only when the UE is both queue-exposed and likely to benefit from waiting longer for a delayed response.

The computation is lightweight. For each registration batch, the method estimates the nearest GS and ISL hop count for each UE, groups UEs by hop count to compute $C(k)$, and then evaluates the closed-form timer expression. The dominant cost is nearest-GS lookup, $O(|\mathcal{U}||\mathcal{G}|)$, while the remaining steps are linear in $|\mathcal{U}|$.

\section{Experimental Results}
In this section, we evaluate the proposed UE-specific NAS timer adaptation method against 3GPP default timers and AstroTimer using NAS registration simulations under different UE loads and packet-loss settings.

\subsection{Experimental Setup}

We use a Python-based simulator of the main NAS registration exchanges between the UE and AMF, including T3510, T3550, and T3560. For satellite NG-RAN access, 3GPP specifies 27~s for T3510 and 11~s for T3550/T3560, these values form the 3GPP baseline.

The simulated NTN--TN network comprises terrestrial UEs, a 14-satellite LEO relay chain, an FL to the gateway, and an AMF in the 5G core. The main experiments set $N_{\mathrm{chain}}=14$ to capture the multi-hop path variation inherent in dense LEO routing and to clearly expose its impact on timer performance. Tests conducted with shorter chains exhibited the same qualitative trends. Each UE is assigned to a serving satellite based on its estimated ISL hop distance to the gateway. All schemes use identical UE placement, routing, loss, AMF service, and energy settings. UEs initiate registration simultaneously and may make up to five attempts, with the configured waiting interval applied after each unsuccessful attempt.










\begin{table}[t]
\centering
\caption{Main simulation and timer-model parameters.}
\label{tab:sim-params}
\small
\setlength{\tabcolsep}{4pt}
\begin{tabular}{lc|lc}
\hline
\textbf{Parameter} & \textbf{Value} &
\textbf{Parameter} & \textbf{Value} \\
\hline
$N_{\mathrm{chain}}$ & 14
& $h$ & $550\,\mathrm{km}$ \\

$N_{\mathrm{orb}}\times N_{\mathrm{spo}}$ & $72\times22$
& $i$ & $53^\circ$ \\

$\theta_{\mathrm{el,min}}$ & $25^\circ$
& $\eta$ & 0.3 \\

$r_{\mathrm{GS}}$ & $1000\,\mathrm{km}$
& $\kappa$ & 1.5 \\

$D_{\mathrm{ISL}}$ & $16\,\mathrm{ms}$
& $D_{\mathrm{FL}}$ & $2\,\mathrm{ms}$ \\

$C_{\mathrm{SL}}$ & $2\,\mathrm{Mbit/s}$
& $C_{\mathrm{ISL}}=C_{\mathrm{FL}}$ & $20\,\mathrm{Gbit/s}$ \\

$\mu_{\mathrm{sat}}$ & $3500\,\mathrm{pkt/ms}$
& $\mu_{\mathrm{NF}}$ & $0.3\,\mathrm{req/ms}$ \\

$\rho_{\mathrm{NF}}$ & 0.8
& $t_{\mathrm{brs}}$ & $1\,\mathrm{ms}$ \\

$\alpha_0$ & 0.5
& $R_{\mathrm{ref}}$ & 5 \\

$A_{\max}$ & 5
& $\Delta T$ & $1\,\mathrm{ms}$ \\

$P_{\mathrm{idle}}$ & $20\,\mathrm{mW}$
& $P_{\mathrm{active}}$ & $\mathcal{U}[50,75]\,\mathrm{mW}$ \\
\hline
\end{tabular}
\end{table}

Table~\ref{tab:sim-params} summarizes the fixed parameters. $N_{\mathrm{orb}}\times N_{\mathrm{spo}}$ gives the Walker-shell geometry used for path estimation, while $\rho_{\mathrm{NF}}$ and $t_{\mathrm{brs}}$ denote the AMF load and burst window. Across experiments, $|\mathcal{U}|\in\{3000,4000,5000\}$ and $p_{\mathrm{loss}}\in\{0,0.1,\ldots,0.5\}$ are varied.

\subsection{Timer Configuration}

We first compare the timer values assigned by the three schemes before running the end-to-end registration procedure. The 3GPP baseline uses fixed NAS timer values specified for satellite access, i.e., T3510 = 27~s and T3550/T3560 = 11~s. AstroTimer computes one closed-form timer value for each NAS timer under a given network load, and the same value is applied to all UEs. In contrast, the proposed method assigns UE-specific values for T3510, T3550, and T3560 according to each UE's path delay and adaptive endpoint weight.

\subsection{Performance Metrics}

We evaluate the three timer schemes using registration length, UE energy consumption, and the number of registration attempts. Registration length measures the time from the first registration request to successful completion. UE energy consumption is accumulated over the registration period based on the UE idle and active power states. The number of registration attempts reflects the signaling overhead caused by timer expiration, packet loss, and repeated NAS retransmissions.

\subsubsection{Registration length}

\begin{figure}[t]
    \centering
    \includegraphics[width=\linewidth]{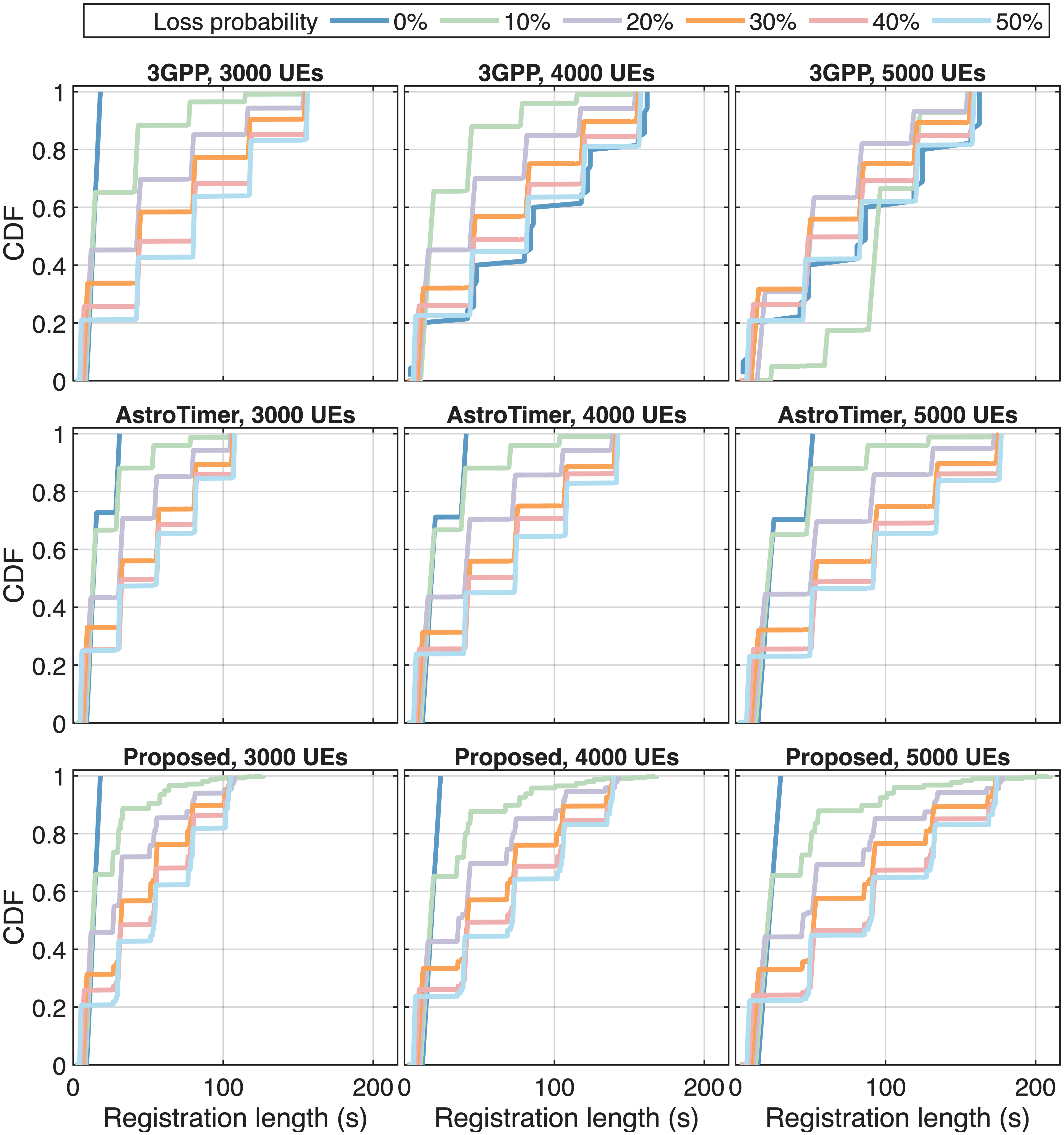}
    \caption{CDF of registration length for 3GPP, AstroTimer, and the proposed method.}
    \label{fig:reg-length-cdf}
\end{figure}

Fig.~\ref{fig:reg-length-cdf} shows the CDF of registration length. In the loss-free setting, the proposed method consistently shifts the CDF to the left compared with AstroTimer. The mean registration length is reduced by 20.5\%, 21.5\%, and 22.0\% for 3000, 4000, and 5000 UEs, respectively. This indicates that UE-specific timer assignment avoids the unnecessary waiting introduced by a single global timer. As packet loss increases, the gap becomes smaller because registration delay is increasingly dominated by failed transmissions and repeated attempts. Under these lossy settings, the proposed method remains comparable to AstroTimer while avoiding the severe degradation observed with fixed 3GPP timers under heavy load.

\begin{figure}[t]
    \centering
    \includegraphics[width=\linewidth]{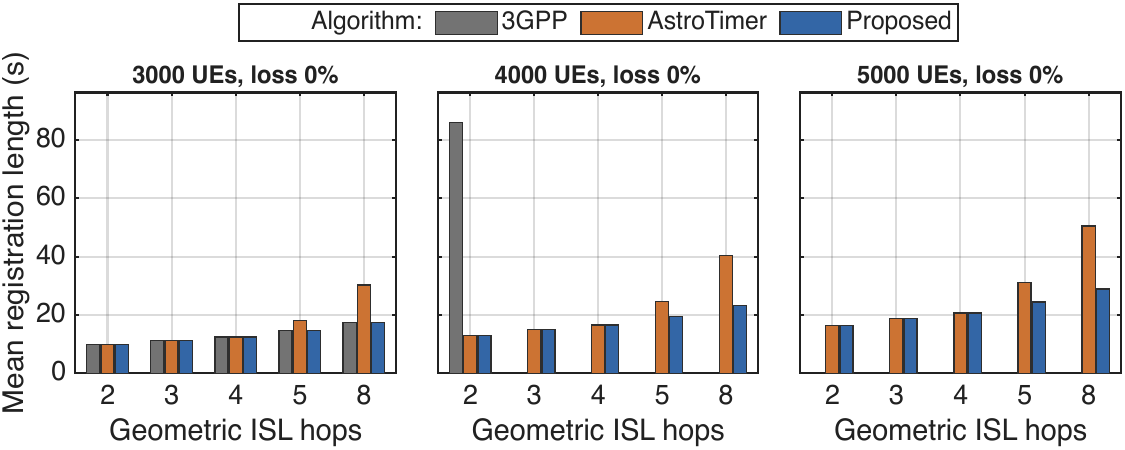}
    \caption{Mean registration length grouped by geometric ISL hop count under loss-free operation.}
    \label{fig:hop-reg-length}
\end{figure}

Fig.~\ref{fig:hop-reg-length} shows the loss-free registration length grouped by geometric ISL hop count. For low-hop UEs (2--4 hops), the proposed method and AstroTimer have the same registration length because both complete registration in one attempt; the benefit of UE-specific timer adaptation becomes more apparent as the ISL hop count increases. At hop 5, the proposed method reduces mean registration length over AstroTimer by 18.9\%, 20.7\%, and 21.8\% for 3000, 4000, and 5000 UEs, respectively. At hop 8, the reductions are 42.5\%, 42.6\%, and 42.7\%. In contrast, 3GPP succeeds only for 45 UEs at 4000 UEs and fails to complete any registration at 5000 UEs.

\subsubsection{UE energy consumption}
\begin{figure}[t]
    \centering
    \includegraphics[width=\linewidth]{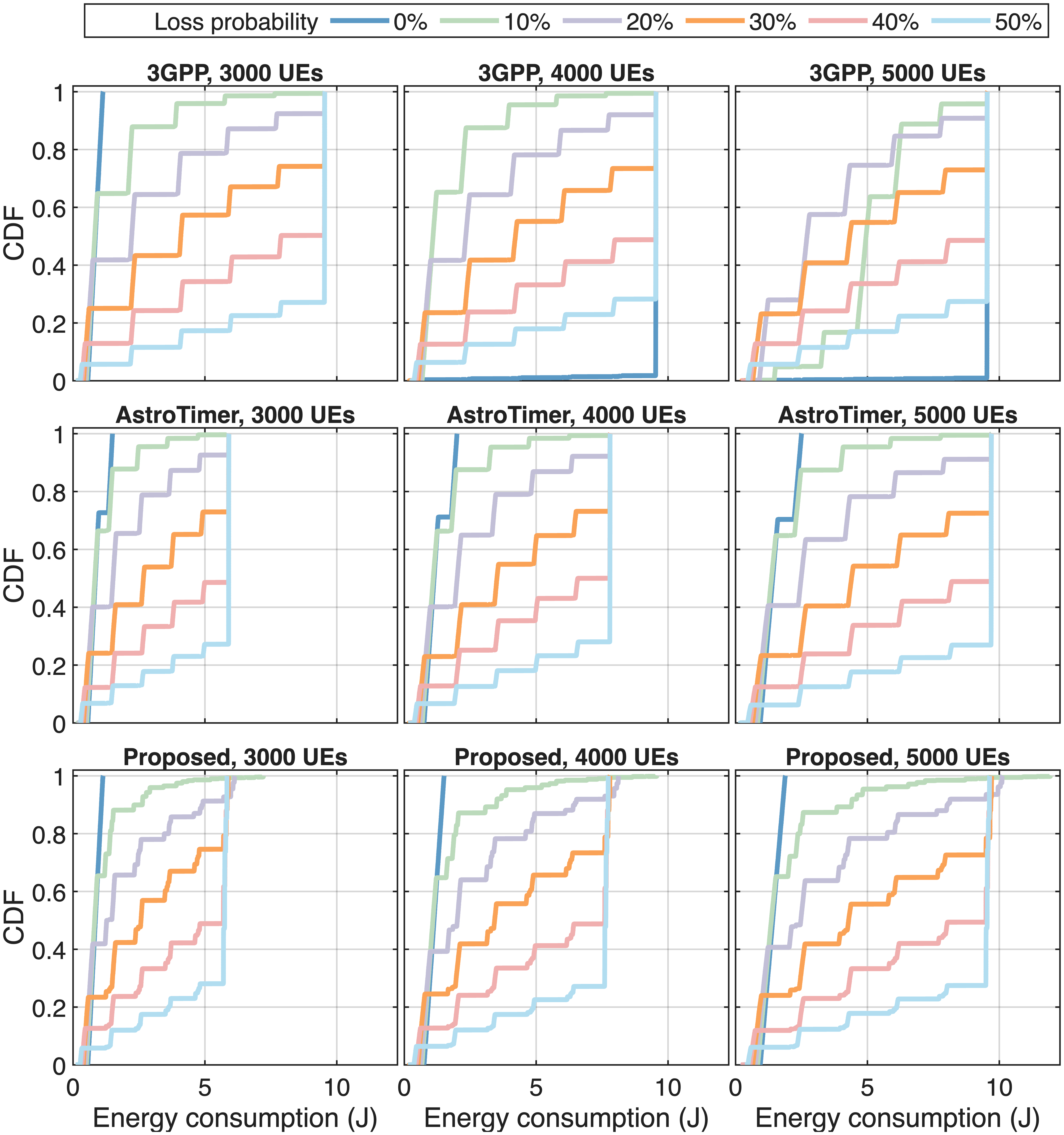}
    \caption{CDF of UE energy consumption for 3GPP, AstroTimer, and the proposed method.}
    \label{fig:energy-cdf}
\end{figure}

Fig.~\ref{fig:energy-cdf} reports the CDF of UE energy consumption. The proposed method reduces mean energy consumption by 10.5\%, 11.2\%, and 11.5\% over AstroTimer in the loss-free setting for 3000, 4000, and 5000 UEs, respectively. The gain comes from reducing over-provisioned waiting time without introducing extra retransmissions. Under packet loss, the energy gap narrows as retransmissions dominate UE active time, but the proposed method remains generally comparable to AstroTimer.

\subsubsection{Registration attempts}
\begin{figure}[t]
    \centering
    \includegraphics[width=\linewidth]{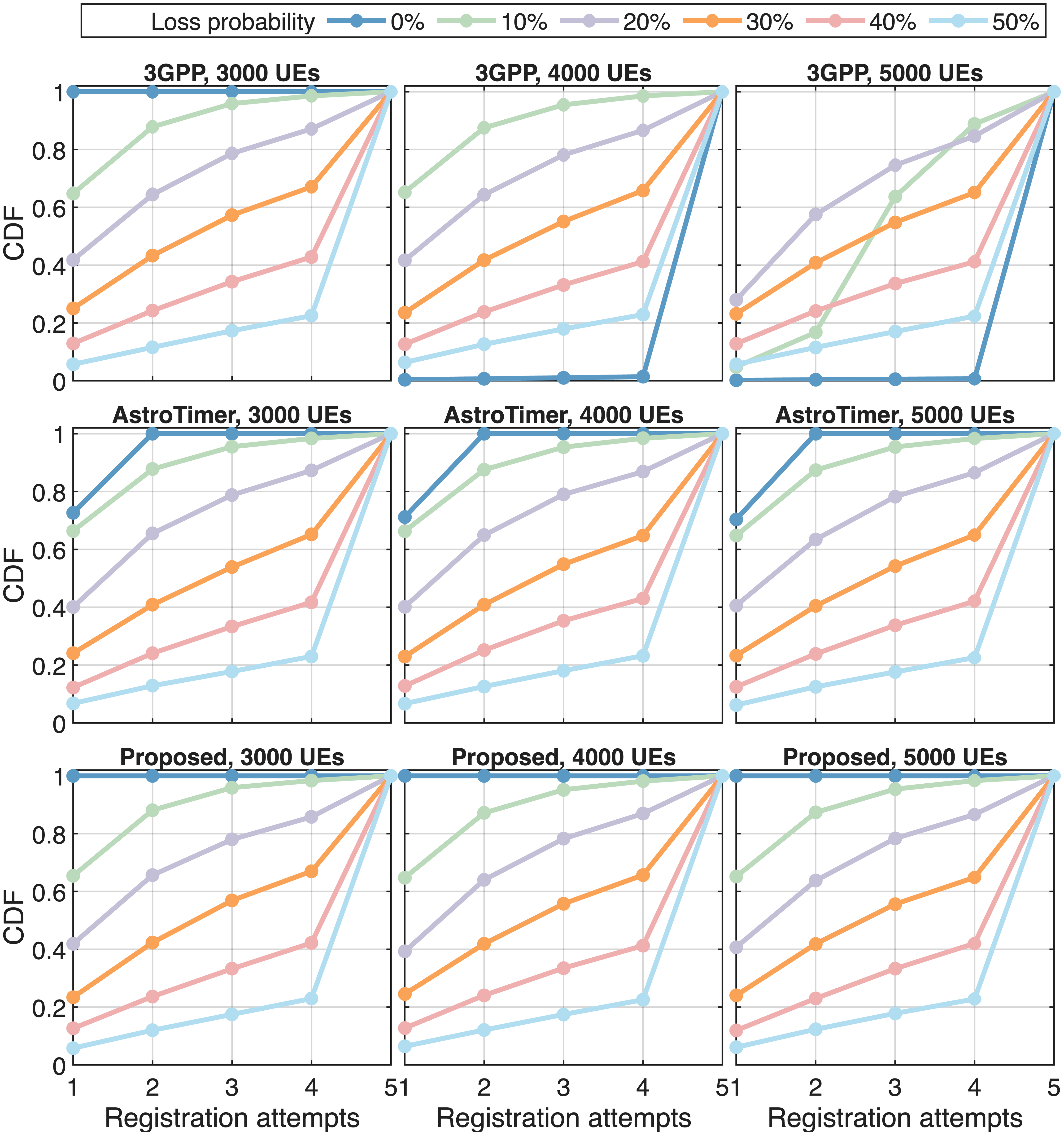}
    \caption{CDF of registration attempts for 3GPP, AstroTimer, and the proposed method.}
    \label{fig:attempts-cdf}
\end{figure}

Fig.~\ref{fig:attempts-cdf} shows the CDF of registration attempts per UE. Without packet loss, the proposed method completes all registrations in a single attempt across all tested UE loads, whereas AstroTimer still causes a fraction of UEs to retransmit. This confirms that per-UE timer sizing reduces avoidable retransmissions caused by path heterogeneity and AMF queueing. At high packet-loss rates, all schemes require more attempts, indicating that retransmissions are then dominated by packet drops rather than timer configuration.

\begin{figure}[t]
    \centering
    \includegraphics[width=\linewidth]{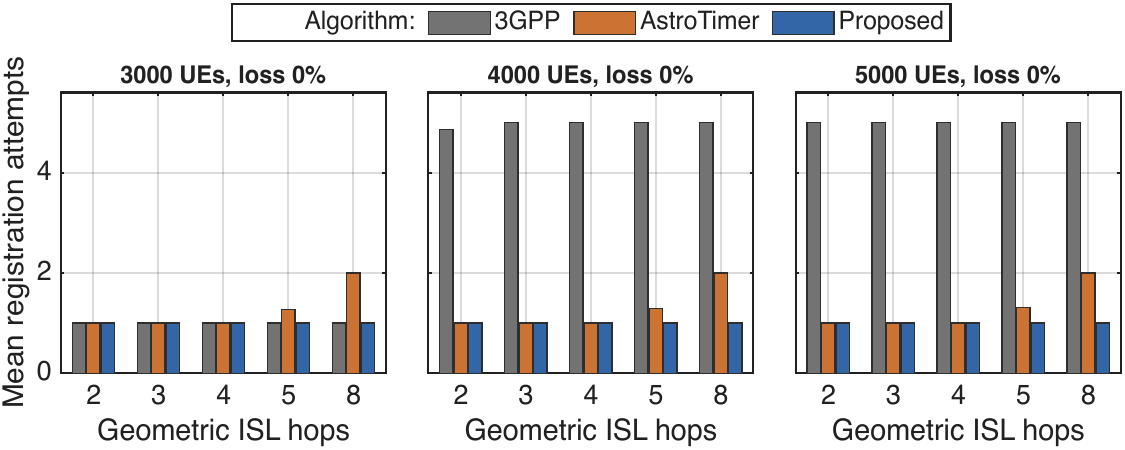}
    \caption{Mean registration attempts grouped by geometric ISL hop count under loss-free operation.}
    \label{fig:hop-attempts}
\end{figure}

Fig.~\ref{fig:hop-attempts} reports the mean registration attempts by geometric ISL hop count under 0\% packet loss. The proposed method completes all UEs in one attempt across all hop groups and UE loads. AstroTimer also needs one attempt for hops 2--4, but requires 1.27--1.32 attempts at hop 5 and 2.00 attempts at hop 8. The 3GPP baseline reaches nearly the retry limit under heavier load: at 4000 UEs it requires 4.87 attempts at hop 2 and 5.00 attempts for hops 3--8, while at 5000 UEs it requires 5.00 attempts for every hop group.

\section{Conclusion}
In this paper, we proposed a UE-specific NAS timer adaptation method for LEO satellite networks. We modeled the timer path component using SL geometry and ISL hop count, and adapted the endpoint-delay component according to each UE's expected AMF queue exposure and path reliability. Through simulation, we showed that our method reduces registration latency, UE energy consumption, and avoidable registration attempts compared with fixed and global timer configurations, particularly when timer over-provisioning is a major source of inefficiency. Future work may explore dynamic satellite mobility, time-varying registration bursts, and online estimation of path reliability in operational NTN deployments.

\ifCLASSOPTIONcaptionsoff
  \newpage
\fi




\bibliographystyle{IEEEtran}
\bibliography{references}

\end{document}